\documentclass[prl,a4paper,aps,twocolumn,showpacs,superscriptaddress,groupedaddress]{revtex4}
\usepackage[dvips]{graphicx}
\usepackage{ae}
\usepackage[T1]{fontenc}
\usepackage[ansinew]{inputenc}
\usepackage{amsmath}
\usepackage{amssymb}
\usepackage{graphicx}
\usepackage{color}
\usepackage[colorlinks]{hyperref}
\usepackage{lscape}
\begin{document}
\title{One Sided indeterminism alone is not a useful resource to simulate any nonlocal correlation}
\author{Biswajit Paul}
\email{biswajitpaul4@gmail.com}
\affiliation{Department of Mathematics, St.Thomas' College of Engineering and Technology,4,Diamond Harbour Road,Alipore, Kolkata-700023, India.}
\author{Kaushiki Mukherjee}
\email{kaushiki_mukherjee@rediffmail.com}
\affiliation{Department of Applied Mathematics, University of Calcutta, 92, A.P.C. Road, Kolkata-700009, India.}

\author{Debasis Sarkar}
\email{dsappmath@caluniv.ac.in}
\affiliation{Department of Applied Mathematics, University of Calcutta, 92, A.P.C. Road, Kolkata-700009, India.}
\begin{abstract}
Determinism, no signaling and measurement independence are some of the constraints required for framing Bell inequality. Any model simulating nonlocal correlations must either individually or jointly give up these constraints. Recently M. J. W. Hall (Phys Review A, \textbf{84}, 022102 (2011)) derived different forms of Bell inequalities under the assumption of individual or joint relaxation of those constraints on both(i.e., two) the sides of a bipartite system. In this work we have investigated whether one sided relaxation can also be a useful resource for simulating nonlocal correlations or not. We have derived Bell-type inequalities under the assumption of joint relaxation of these constraints only by one party of a bipartite system. Interestingly we found that any amount of randomness in correlations of one party in absence of signaling between two parties is incapable of showing any sort of Bell-CHSH violation whereas signaling and measurement dependence individually can simulate any nonlocal correlations. We have also completed the proof of a recent conjecture due to Hall (Phys. Rev. A \textbf{82}, 062117 (2010); Phys. Rev. A \textbf{84}, 022102 (2011)) for one sided relaxation scenario only.
\end{abstract}
\date{\today}
\maketitle

{\bf Introduction:}
Bell inequalities are certain constraints on correlations between space-like separated measurements which are satisfied in any local realistic theory \cite{Bell}. The inequalities are violated by quantum predictions for some entangled states. The set of assumptions made in the derivation of Bell inequalities are realism, the experimenter’s freedom to choose the measurement settings and two locality conditions: (i) outcome independence and (ii) parameter independence \cite{Jar,See};  parameter independence is also known as no signaling condition. Violation of Bell inequalities by any physical theory thereby give rise to some queries:\textit{ are the predictions made by the theory incorrect? or, whether at least one of these applied postulates incompatible with the description of the natural phenomena?} As quantum mechanical predictions tally with the experimental data so only the second query is relevant in this regard. Recently lot of work has been done to simulate nonlocal correlations by violating these assumptions jointly or individually \cite{BRan,Gisin,Hall1,Hall2,Hall3,Manik,Kar,Paw,Bis,Manik1}. In particular, simulation protocols for singlet state correlation was given in \cite{Mas,Ton,Deg,Cer}. The main objective of all these works (\cite{BRan,Gisin,Hall1,Hall2,Hall3,Kar,Paw,Bis,Manik1,Manik2}) was to simulate nonlocal correlations assuming two sided relaxation of these constraints collectively or individually. The question that naturally arises in this context is whether joint or individual one sided relaxation of these constraints can achieve any Bell violation. If yes, then  how much of those constraints has to be relaxed on one side? i.e., our motivation in this paper is asymmetric characterization of quantum correlation through relaxation of constraints. Such a scenario is possible if one party performs measurement before the other one. Interestingly we find that no Bell violation is possible only by one sided relaxing of determinism, i.e., at least some amount of signaling should be sent from one party to the other or atleast some amount of measurement independence is to be relaxed on one side or both(relaxing no signaling and measurement independence simultaneously) unlike that in two sided relaxation scenario where  nonlocal correlations can be simulated by relaxing determinism \cite{Hall2,Hall3} only. On the contrary, one sided signaling and/or is/are useful resource(s) for simulation of nonlocal correlations. In this regard, we obtain Bell type inequalities by relaxing constraints.

Now consider a joint experiment between two parties, Alice and Bob. Suppose each party has dichotomic measurement settings with inputs $x$ and $y$ for Alice and Bob respectively. Let $a$ and $b$ $\in\{-1,\,1\}$ label the possible outcomes of Alice and Bob respectively. The results of this experiment can be expressed in terms of the statistical correlations $p(a,\,b|\,x,\,y)$. Now, for correlations corresponding to any experiment, there always exists an underlying model that depends upon a variable $\lambda$ (say), frequently referred to as hidden variable. Thus the correlations thereby depend on this underlying hidden variable $\lambda$ and by Bayes' theorem one has :
\begin{equation}\label{p1}
    p(a,\,b|x,\,y)\,=\,\int d\lambda \, p(a,\,b|x,\,y,\,\lambda)\,p(\lambda|x,\,y).
\end{equation}
 For discrete range of values of $\lambda$ integration will be replaced by summation. Before discussing our main results, we briefly describe the notions of some constraints required for our further discussions. \\
No signaling is the constraint that the underlying marginal distributions associated with the setting of one party does not depend on that with the setting of the other party, i.e., if

\hspace{2cm}$p^{(1)}(a|x,y,\lambda)= p^{(1)}(a|x,y',\lambda),$

\begin{equation}\label{p2}
            p^{(2)}(b|x,y,\lambda)= p^{(2)}(b|x',y,\lambda)
\end{equation}
hold for all pairs $(x,\,y)$, $(x,\,y')$ and $(x',\,y)$ of the model.
The degree of signaling is then defined as the maximum shift possible in an underlying marginal probability distribution for one party, due to the alteration of measurement settings by the other. It is quantified by the variational distance \cite{Hall2,Hall3}
\begin{equation}\label{p3}
    S_{1\rightarrow2}\,:=\,\sup_{x,x',y,b,\lambda}|\,p^{(2)}(b|x,\,y,\,\lambda)\,-\,p^{(2)}(b|x',\,y,\,\lambda)\,|
\end{equation}
\begin{equation}\label{p4}
    S_{2\rightarrow1}\,:=\,\sup_{x,y,y',a,\lambda}|\,p^{(1)}(a|x,\,y,\,\lambda)\,-\,p^{(1)}(a|x,\,y',\,\lambda)\,|
\end{equation}
where $a$, $b$, $x, x', y$ and $y'$ have their usual meanings. According to the definition, $ S_{1\rightarrow2}$ is the maximum possible deviation in an underlying marginal probability distribution for the second observer, induced via the change of a measurement setting of the first observer. Given $\lambda$, if $ S_{1\rightarrow2}\,>\,0$, then the first observer in principle can send a signal to the second observer by merely altering its own measurement setting. Hence if $ S_{1\rightarrow2}\,>\,0$ and $S_{2\rightarrow1}\,=\,0$ then no signaling is relaxed only on second observer's side. This is referred to as `one sided signaling'. \\
The overall degree of signaling, for a given underlying model is defined by,
\begin{equation}\label{p5}
    S:=\,max\,\{S_{1\rightarrow2},\,S_{2\rightarrow1}\}.
\end{equation}
Thus, 0$\leq S \leq $1, where two extreme values $0$ and $1$ represent no signaling and $100\%$ signaling or simply signaling, respectively.\\
In a deterministic correlation model all the outcomes being predictable with certainty for any given knowledge of $\lambda$, correlation terms are either 0 or 1, i.e., $ p(a,\,b|x,\,y,\,\lambda)\,\in\{0,\,1\}$. The degree of indeterminism of an underlying model may be defined as the measure of deviation of the marginal probabilities from the deterministic values $0$ and $1$. The local degree of indeterminism $I_j$ may be defined \cite{Hall2,Hall3} as the smallest positive number, such that the corresponding marginal probabilities lie in $[0,\,I_j]\,\bigcup\,[1-I_j,1]$,
\begin{equation}\label{p6}
    I_j\,:=\,\sup_{\{x,\,y,\,\lambda\}}\,\min_z\{p^{j}(z|x,\,y,\,\lambda),\,1-\,p^{j}(z|x,\,y,\,\lambda)\}.
\end{equation}

Hence $I_j\,=$ 0 if and only if the corresponding marginal is deterministic.\\
The overall degrees of indeterminism  for the model may be defined as,
 \begin{equation}\label{p7}
     I\,:=\,max\{I_1,\,I_2\}.
 \end{equation}
Thus, 0$\leq I \leq $1/2, where two extreme values $0$ and $\frac{1}{2}$ represent determinism and indeterminism, respectively.\\
\textit{Complementary Relation between Signaling and Indeterminism}: Without loss of generality let Alice sends signal to bob. Due to signaling, any deviation in a marginal probability value must either retain the value in the same subinterval $[0,\,I_2]$ (or $[1-I_2,1]$) for $S_{1\rightarrow2} < 1-2I_2$ or, shift it across the gap between the subintervals $(S_{1\rightarrow2}\geq 1-2I_2)$. Clearly, for the region $S_{1\rightarrow2}<1-2I_2$, $S_{1\rightarrow2} \leq I_2$. This provides us the relation: $I_2\geq \textmd{min}\{S_{1\rightarrow2},\,(1-S_{1\rightarrow2})/2\}$.\\
As noted in introduction any deterministic and no signaling model must satisfy Bell Inequalities if experimenters enjoy complete measurement independence. In order to generate models violating that inequality, the properties of no signaling and that of determinism are relaxed to some extent, thereby introducing signaling and indeterminism in the models. The extent of relaxation of these two constraints on both sides can be quantified with the help of the corresponding relaxed Bell-type inequality \cite{Hall2,Hall3,Bis}. Relaxation of these two constraints is not the only way to simulate nonlocal correlations. Measurement independence can also be relaxed. Measurement independence is the constraint that
the distribution of the underlying variable is independent of the measurement settings, i.e.,
\begin{equation}\label{p8}
 p(\lambda|x,\,y)\,=\, p(\lambda|x',\,y').
\end{equation}
Thus, \textit{Measurement dependence(M)} may be interpreted as a measure to quantify the degree of violation of measurement independence by the underlying model. It is defined as \cite{Hall1}:
\begin{equation}\label{p9}
    M:=\,\sup_{x,x',y,y'}\int d\lambda\,|p(\lambda|x,\,y)\,-\,p(\lambda|x',\,y')|.
\end{equation}
Therefore, if Eq.(\ref{p8}) holds, then $M\,=\,0$. On the contrary, the maximum possible value of $M$ is given by $M_{max}\,=\,2$ implying complete measurement dependence in this case. \\
The fraction of measurement independence corresponding to a given model is defined by \cite{Hall1},
\begin{equation}\label{p9}
    F:=\,1\,-\,M/2.
\end{equation}
Thus $0\,\leq F \leq\,1$, with $F\,=\,0$ for the case where $M\,=\,2$. Geometrically, F represents the minimum degree of overlap
between any two underlying distributions $p(\lambda|x,\,y)$ and $p(\lambda|x',\,y')$.\\
Also local degrees of measurement dependence are defined analogously \cite{Hall1,Hall3},
\begin{eqnarray}\label{p10}
  M_1:&=& \,\sup_{x,x',y}\int d\lambda\,|p(\lambda|x,\,y)\,-\,p(\lambda|x',\,y)|;\\
 M_2:&=& \,\sup_{x,y,y'}\int d\lambda\,|p(\lambda|x,\,y)\,-\,p(\lambda|x,\,y')|.
\end{eqnarray}
 \\
 Below we describe a model where one sided signaling indeterminism and measurement dependence are considered.\\
 \textit{Relaxed Bell inequalities:} In a system of two parties (Alice and Bob), it is assumed that Bob sends no signal to Alice but a signal is sent from Alice to Bob i.e.,no signaling is preserved by the correlations shown by Alice's measurement but is relaxed on Bob's side. Hence Eq.(\ref{p3}) and Eq.(\ref{p4}) imply $S_{1\rightarrow2} > 0$ and $S_{2\rightarrow1}= 0.$ For this we assume Alice's  measurements are in the past of Bob's measurements. It is also assumed that determinism and measurement independence are relaxed simultaneously only on Bob's side. Since relaxation of no signaling, determinism and measurement independence are relaxed jointly or individually only on Bob's side so we refer this scenario as `one sided relaxation scenario'. The extent of minimum possible relaxation in this context is given by the following theorem.\\

\textit{Theorem}: Suppose $x,x'$ and $y,y'$ be the measurement settings for Alice and Bob respectively and the measurement outcomes for each party be 1 or -1. If $\langle xy \rangle$ denotes the average product of the measurement outcomes for joint measurement settings $x$ and $y,$ then
\begin{equation}\label{B1}
    \langle xy \rangle + \langle xy' \rangle + \langle x'y \rangle - \langle x'y' \rangle \leq B(I_2, S_{1\rightarrow 2} , M_2)
\end{equation}
where $I_2$, $S_{1\rightarrow 2}$ and $M_2$ are the values of indeterminism, signaling and measurement dependence respectively on Bob's side, for any underlying models with
\begin{equation}\label{B2}
\begin{split}
    B(I_2, S_{1\rightarrow 2}, M_2) &= 4 - (1 - I_2)(2 - M_2),\,\textmd{for} \, S_{1\rightarrow 2} < 1 - 2I_2 \\
    &\hspace{2cm}\textmd{and} \, M_2 < 2 \,(\text{tight upper bound})\\
    &= 4 - (1 - S_{1\rightarrow2})(2 - M_2),\,\textmd{for} \, S_{1\rightarrow 2} \geq 1 - 2I_2\\
    &\hspace{2cm}\textmd{and} \, M_2 < 2 \,(\text{tight upper bound})\\
                &= 4,                    \qquad \qquad \qquad \qquad \qquad \textmd{otherwise}.
\end{split}
\end{equation}

where $\langle xy \rangle$ is the average product of the measurement
outcomes for joint measurement settings.
Since $B(0,0,0)=2$, the inequality reduces to well known Bell-CHSH  inequality \cite{Cla} for models satisfying no signaling, determinism and measurement independence. Suppose, the Bell-CHSH inequality be violated by an amount $V$. Hence the corresponding model must satisfy the relation
\begin{equation}\label{p13}
    B(I_2,\,S_{1\rightarrow 2},M_2)\,\geq\,2\,+\,V.
\end{equation}
Measurement independent model can be obtained from the above theorem
if we consider $M_2 = 0$ . Hence the bounds get modified as:
\begin{equation}\label{p12}
   \begin{split}
     B(I_2,\,S_{1\rightarrow2},0) &=\, 2\,+\,2I_2\qquad\qquad  \textmd{for}\quad S_{1\rightarrow2}<\,1\,-\,2I_2 \,\\
       & =\,2+2S_{1\rightarrow2}\qquad \textmd{for}\quad S_{1\rightarrow2}\geq\,1\,-\,2I_2.
   \end{split}
\end{equation}
From the proof of the theorem, it is seen that for $ S_{1\rightarrow2}<\,1\,-\,2I_2 $, the upper bound of $B(I_2,S_{1\rightarrow2},0)$ is achieved when  $S_{1\rightarrow2} = I_2$. Now when $S_{1\rightarrow2} =I_2$ then $I_2<\frac{1}{3}$. Thus,  $B(I_2,S_{1\rightarrow2},0)$ can reach a maximum value  less than $2.67$. So maximum quantum violation, hence simulation of singlet correlation is impossible  for $S_{1\rightarrow2}<1-2I_2$. This proves the conjecture mentioned in \cite{Hall2,Hall3} completely  for one sided relaxation of constraints only and we can conclude from the above that for $S_{1\rightarrow2}<1-2I_2$, it is not possible to simulate any nonlocal correlations by introducing only one sided indeterminism. It follows via Eq. (\ref{p13}) that for $S_{1\rightarrow2}<1-2I_2$, any measurement independent model must assign $I_2\geq\,\frac{V}{2} $ and $S_{1\rightarrow2}\geq\,\frac{V}{2}$ for $V$($V<0.67$) amount of violation($I_2<1/3$ and $S_{1\rightarrow2}<1/3$ as said earlier). Model saturating this bound is given in the Appendix A.\\
\begin{figure}[htb]

\includegraphics[width=2.5in]{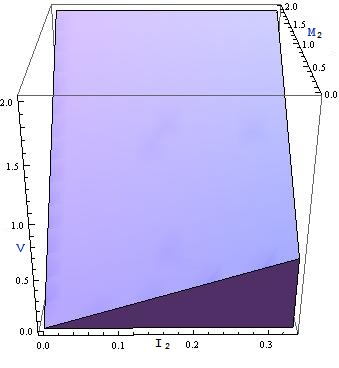}
\caption{\emph{The tradeoff between $I_2$ and $M_2$ for any amount of V is plotted when  $S_{1\rightarrow2}\,<\,1-2I_2$. Clearly, the maximum value that the upper bound $B(I_2,\,S_{1\rightarrow2},0)$ } can achieve is $2.67$ for $S_{1\rightarrow2}\,<\,1-2I_2$.}
\end{figure}\\
 A further consequence of the theorem is that violation from $2.67$ to $4$ can be achieved only when $S_{1\rightarrow2} \geq 1-2I_2$. For $S_{1\rightarrow2}\geq 1-2I_2$ it follows from Eq. (\ref{p13}) that for $V$ ($0<V\leq2$) amount of violation,  corresponding measurement model must assign atleast $V/2$ amount of signaling (hence atleast $\frac{1-S_{1\rightarrow2}}{2}$ amount of indeterminism).  In particular, for generation of singlet state correlation, i.e., for $V=2\sqrt{2}-2$ \cite{Ts} atleast $41\%$ of signaling(hence $59\%$ of indeterminism) is required. It is seen from (\ref{p12}), for  $S_{1\rightarrow2}\,=\,1$,  $B(I_2,\,S_{1\rightarrow2},0)\,=\,4$. Thus maximal violation can be achieved only when $100\%$ communication of signaling takes place from Alice to Bob. Model saturating this bound is given in the Appendix A. For $S_{1\rightarrow2}< 1-2I_2$ case, the upper bound $2+2I_2$ is achieved for $S_{1\rightarrow2} = I_2$. Hence if $S_{1\rightarrow2}=0,$ then no violation can be obtained.  Similarly for $S_{1\rightarrow2}\geq1-2I_2$ case the upper bound $B(I_2, S_{1\rightarrow 2},0)$ is $2+2S_{1\rightarrow2}$. Hence if $S_{1\rightarrow2}=0$ then no violation can be obtained. This in turn points out the fact that complete randomness of outputs of only one party in absence of any sort of communication between the parties is insufficient to  simulate any nonlocal correlation contrary to two sided relaxation scenario where  $B(I,0, 0)$ model exists\cite{Hall2,Hall3,Cer}.\\
Now we consider $B(I_2,S_{1\rightarrow2},M_2)$ model for $M_2\neq0$.
The corresponding model must satisfy the relation
\begin{equation}\label{B3}
    B(I_2,\,S_{1\rightarrow2},\,M_2)\,\geq\,2\,+\,V.
\end{equation}
The upper bound in Eq.(\ref{B2}) for $S_{1\rightarrow2}< 1 - 2I_2$ and $M_2 < 2$ is obtained when $S_{1\rightarrow2} = I_2$. Thus, a one sided signaling, indeterministic and measurement dependent model exists for any violation $V < 2$ and for $S_{1\rightarrow2}< 1 - 2I_2$, $M_2 < 2$ if and if only if it satisfy the following condition:
\begin{equation}\label{B3}
 I_2<\frac{1}{3}, \,S_{1\rightarrow2}<\frac{1}{3} \quad \textmd{and}\quad 2I_2 + M_2(1-I_2)\geq V.
\end{equation}
Now for any amount of violation $V$ the tradeoff between $M_2$ and $I_2$ is shown in the FIG.1 (as in $S_{1\rightarrow2}<1-2I_2$ case, the bound is obtained for $S_{1\rightarrow2}=I_2,$ hence this is also a tradeoff between $S_{1\rightarrow2}$ and $M_2$). \\
From Eq.(\ref{p13}) and Eq.(\ref{B2}) it is clear that the inequality $2S_{1\rightarrow2} + M_2(1-S_{1\rightarrow2})\geq V$ gives the tradeoff between $S_{1\rightarrow2}$ and $M_2$ if $S_{1\rightarrow2}\geq 1 - 2I_2$ and $M_2 < 2$. Clearly from Eq.(\ref{B2}) if $S_{1\rightarrow2}\,=\,1$, and/or $M_2 = 2$,\, $B(I_2,\,S_{1\rightarrow 2},M_2)\,=\,4$.  Thus, maximum violation can also be reached when $100\%$ of communication of
signaling takes place from Alice to Bob and/or there is complete measurement dependence for Bob(irrespective of amount of indeterminism on Bob's side).\\
\begin{figure}[htb]

\includegraphics[width=2.5in]{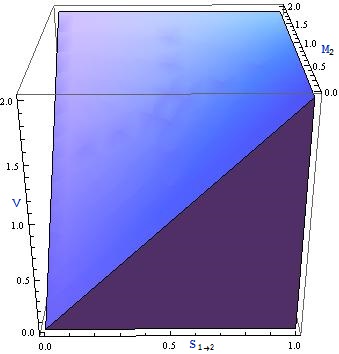}
\caption{\emph{The figure represents the tradeoff between signaling, indeterminism and the amount of violation for  $S_{1\rightarrow2}\geq 1 - 2I_2$. In particular, region lying on the $V$$S_{1\rightarrow2}$  plane(i.e $M_2 = 0$ plane) represents the amount of signaling required for V amount of violation of Bell-CHSH inequality.} }
\end{figure}\\
\textit{Local deterministic model:} In this case, both $I_2 = 0 $ and $S_{1\rightarrow2}=0$. So, Equation (\ref{B2}) reduces to,
 \begin{equation}\label{B4}
  B(0,\,0,\,M_2) = min \{2 + M_2, 4\}.
 \end{equation}
Hence from Eq. (\ref{B2}), a local deterministic model exists for simulating a singlet state correlation if and only if $M \geq V = 2\sqrt{2} - 2 \approx 0.82 $. So $59\%$ measurement independence is optimal for simulating singlet correlation when measurement dependency is allowed only on one side \cite{Manik}. So, our relaxed Bell inequality(Eq. (\ref{B1}) and (\ref{B2})) gives a general result from which results of \cite{Manik} can be obtained as a particular case.\\
\begin{figure}[htb]

\includegraphics[width=2.5in]{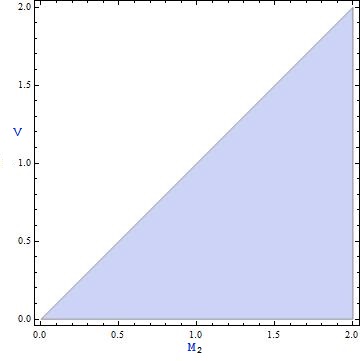}
\caption{\emph{The figure shows the amount of measurement dependence required on bob's side to simulate any nonlocal correlation for an one sided local deterministic model.} }
\end{figure}\\

For $S_{1\rightarrow2}<1-2I_2$ and $M_2 < 2$ case if one allows maximum amount of signaling, i.e., $S_{1\rightarrow2}=1/3$ and maximum amount of indeterminism, i.e., $I_2=1/3$ then atleast $12\%$ of measurement dependence is required.\\
\begin{figure}[htb]

\includegraphics[width=2.5in]{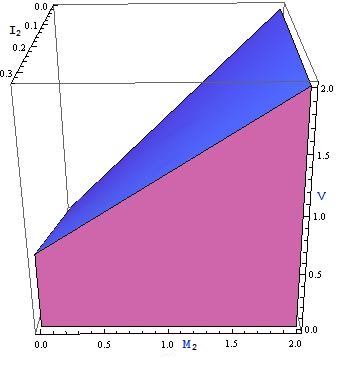}
\caption{\emph{For $S_{1\rightarrow2}< 1 - 2I_2$ case when $I_2=1/3$ then the minimum amount of measurement dependence required to simulate any amount of violation($V$) is given by the $I=\frac{1}{3}$ plane. } }
\end{figure}\\
\textit{Conclusion}: Thus our work provides the condition which every `one sided relaxation model' has to satisfy in order to allow an amount of violation of the Bell-CHSH inequality. From the above discussions, it can now be safely concluded that one sided indeterminism together with one sided signaling can obtain any Bell-CHSH violation but one sided indeterminism alone cannot achieve any sort of Bell-CHSH violation unlike that of individual relaxation of no signaling on one side. This is also the case for individual relaxation of measurement independence on one side as it alone can simulate any amount of Bell-CHSH violation. Besides, joint relaxation of determinism, no signaling and measurement independence on one side can also achieve Bell-CHSH violation. It is thus clear that  individual relaxation of both one sided no signaling and measurement independence are useful resource for any amount of Bell-CHSH violation.

{\bf Acknowledgement:} This work is supported by University Grants Commission(UGC), New Delhi.

\section{Appendix}
{\bf Appendix A: Proof of the Theorem for $M_2 = 0 $.}
The technique of the proof is similar to that given in \cite{Hall2}. The relaxed Bell inequality given in Eq.(\ref{B1}) and Eq.(\ref{B2}) for $M_2 = 0 $ is proved here.
The outcomes of Alice and Bob are labeled by $\pm1$. Let the joint measurement outcomes be $(+,\,+)$, $(+,\,-)$, $(-,\,+)$ and $(-,\,-)$.\\
Let $p(+,\,+|x,\,y,\,\lambda)\,=\,c$; $p^{(1)}(+|x,\,y,\,\lambda)\,=\,m$ and $p^{(2)}(+|x,\,y,\,\lambda)\,=\,n$. Hence if $\langle xy \rangle_{\lambda}$ denotes the average product of the measurement outcomes for a fixed value of $\lambda$, then $\langle xy \rangle_{\lambda}\,=\,1+4c-2(m+n)$ \cite{Hall2}. For positivity of joint probabilities, $\max \{0,\,m+n-1\}\,\leq c\,\leq\,\min\{m,\,n\}$. Thus,\\
\begin{equation}\label{D1}
   2|m+n-1|\,-\,1\,\leq\,\langle xy \rangle_{\lambda}\,\leq\,1-2|m-n|.
\end{equation}
Suitable choices of $c$ give upper and lower bounds.\\
Let, $p_1\,\equiv\,p(*|x,\,y)$, $p_2\,\equiv\,p(*|x,\,y')$, $p_3\,\equiv\,p(*|x',\,y)$ and $p_4\,\equiv\,p(*|x',\,y')$;
$$E({\lambda}):=\,\langle xy \rangle_{\lambda}\,+\,\langle xy' \rangle_{\lambda}\,+\,\langle x'y\rangle_{\lambda}\,-\,\langle x'y' \rangle_{\lambda},$$ then,
\begin{equation}\label{a1}
    E({\lambda})\,\leq\,4\,-\,2J,
\end{equation}
where
\begin{equation}\label{a2}
 J=\,|m_1\,-\,n_1|\,+\,|m_2\,-\,n_2|\,+\,|m_3\,-\,n_3|\,+\,|m_4\,+\,n_4\,-\,1|.
\end{equation}
For suitable choices of $c_1$, $c_2$, $c_3$ and $c_4$ the upper bound is attainable.\\
As Alice's correlations abide by no signaling principle, $m_1\,=\,m_2$ and $m_3\,=\,m_4$. Therefore,
\begin{equation}\label{a3}
 J=\,|m_1\,-\,n_1|\,+\,|m_1\,-\,n_2|\,+\,|m_3\,-\,n_3|\,+\,|m_3\,+\,n_4\,-\,1|.
\end{equation}
Due to determinism on Alice's part, $m_1,\,m_3\in\{0,\,1\}$. The indeterminism and signaling constraints of the theorem on Bob's outcomes imply $n_j\in[0,\,I]\,\bigcup\,[1-I,\,1]\quad \forall \,\textmd{j}=1,2,3,4$ and $|n_1\,-\,n_3|\,\leq\,S$; $|n_2\,-\,n_4|\,\leq\,S$.\\
To maximize $E({\lambda})$, $J$ must be minimized.
There are four possible cases corresponding to the four possible values of $(m_1,\,m_3)$: $(0$,\,0)$, $(1$,\,0)$, $(0$,\,1)$ \,$\text{and}$\, $(1$,\,1)$.
Now for the first case, i.e., for $m_1=0$ and $m_3 = 0$, $J$ defined in Eq.(\ref{a3}) becomes
\begin{equation}\label{a4}
 J=\,|n_1|\,+\,|n_2|\,+\,|n_3|\,+\,|1\,-\,n_4|.
\end{equation}
Since $n_1$,$n_2$,$n_3$ and $(1-n_4)$ all are positive, hence
\begin{equation}
\begin{split}
J  &=\,n_1\,+\,n_3\,+\,n_2\,-\,n_4\,+\,1\\
&\geq  1-S_{1\rightarrow2}+n_1+n_3\\
&\geq  1-S_{1\rightarrow2}\\
\end{split}
\end{equation}
where we have used the constraints $-S_{1\rightarrow2}\leq n_2-n_4 \leq S_{1\rightarrow2}$. Therefore, $J\geq 1-S_{1\rightarrow2}$. Equality is obtained when $n_1 = 0$,$n_3 =0 $ and $n_2-n_4 = -S_{1\rightarrow2}$. Similarly, for the other three cases $(m_1,\,m_3)$: $(1$,\,0)$, $(0$,\,1)$\text{and}$\, $(1$,\,1)$, $J\geq 1-S_{1\rightarrow2}$ where the equality is obtained for the three cases when $\{n_1 = 0$,$n_3 =0 $ and $n_2-n_4 = -S_{1\rightarrow2}\}$, $\{n_2 = 1$,$n_4 = 1 $ and $n_3-n_1 = -S_{1\rightarrow2}\}$ and \{$n_1 = 0$,$n_3 =1 $ and $n_4-n_2 = -S_{1\rightarrow2}\}$ respectively.
Hence in any case
\begin{equation}\label{a5}
 J\geq 1-S_{1\rightarrow2}.
\end{equation}
Now, let $S_{1\rightarrow2}<1-2I_2$, then the maximum value of $S_{1\rightarrow2}$ will give the minimum value of the lower bound of $J$ in the Eq.(\ref{a5}). The maximum value that  $S_{1\rightarrow2}$  can take for $S_{1\rightarrow2}<1-2I_2$ is $I_2$(by definition). Hence for $S_{1\rightarrow2}<1-2I_2$ Eq.(\ref{a5}) gets modified as
\begin{equation}\label{a6}
   J\geq 1-I_2
\end{equation}
and equality is obtained when $S_{1\rightarrow2} = I_2$. The above relation implies (Via Eq.(\ref{a1})) the tight bound $E({\lambda})\,\leq\,2\,+\,2I_2,$ where $E({\lambda})$ achieve maximum value $2\,+\,2I_2,$ for the following cases $\{m_j = 0, n_1=0 , n_3=0, n_4=I,n_2 = 0 (j=1,2,3,4))\}$,$\{m_1 = m_2 = 0, m_3 = m_4 = 1, n_2 = 0 , n_4 = 0, n_3 = I, n_1 = 0\}$, $\{m_1 = m_2 = 1, m_3 = m_4 = 0, n_2 = 1 , n_4 = 1, n_3 = I, n_1 = 0\}$ and $\{m_j= 1, n_1 = 1 , n_3 = 1, n_2 = I, n_4 = 0(j=1,2,3,4)\}$. Finally, let $S_{1\rightarrow2} \geq 1-2I_2$, then the value of at least one pair of marginal probabilities (($n_1$ , $n_3$) and/or ($n_2$ , $n_4$)) must shift across the gap between the subintervals $[0,I]$ and $[1-I,1]$. Now, the constraint $S_{1\rightarrow2} \geq 1-2I_2,$  will not improve the the lower bound $1-S_{1\rightarrow2}$ of $J$ given in Eq.(\ref{a5}), hence $J\geq 1-S_{1\rightarrow2}$, implying (Via Eq.(\ref{a1})) the tight bound $E({\lambda})\,\leq\,2\,+\,2S_{1\rightarrow2},$.

{\bf Appendix. B: Proof of Theorem for $M_2 \neq 0$.}

We start the derivation of the bounds in the theorem ( given in (\ref{B1}) and (\ref{B2})) by defining
\begin{equation}\label{B3}
\begin{split}
 T(\lambda) &= P_{xy}(\lambda)\langle xy \rangle + P_{xy'}(\lambda)\langle xy' \rangle \\
 &\hspace{1cm}+ P_{x'y}(\lambda)\langle x'y\rangle - P_{x'y'}(\lambda)\langle x'y'\rangle
 \end{split}
\end{equation}
where $p_{xy} = p(a,b|x,y,\lambda)$, $p_{xy'} = p(a,b|x,y',\lambda)$, etc., and $P_{xy} = p(\lambda|x,y)$, $P_{xy'} = p(\lambda|x,y')$, etc.\\
Now by Eq. (\ref{D1}) the above relation takes the form
\begin{equation}\label{B4}
    T(\lambda)\leq P_{xy}(\lambda) + P_{xy'}(\lambda) + P_{x'y}(\lambda) + P_{x'y'}(\lambda) - 2J(\lambda)
\end{equation}
where
\begin{equation}\label{B5}
\begin{split}
    J(\lambda)&= P_{xy}(\lambda)|m_{1} - n_{1}| + P_{xy'}(\lambda) |m_2 - n_2| \\
    &+ P_{x'y}(\lambda)|m_3 - n_3| + P_{x'y'}(\lambda)|m_4 + n_4 -1|.
 \end{split}
\end{equation}
Here $m_j$, $n_j$ and $c_j$ all depend on the underlying variable $\lambda$.\\
From the statement of the theorem it is clear that the restrictions are,
\begin{equation}\label{B6}
   m_j\in\{0,1\} \text{,} \,\, n_j \in [0\,,\,I_2] \cup [1-I_2\,,\,1]
\end{equation}
 \begin{equation}\label{B7}
   m_1 = m_2,\, m_3 = m_4
 \end{equation}
 \begin{equation}\label{B8}
   |n_1 - n_3|  \text{,}\,  |n_2 - n_4| \leq S_{1\rightarrow2}
 \end{equation}
 and
 \begin{equation}\label{B9}
 \begin{split}
   \int |P_{xy}(\lambda) - P_{xy'}(\lambda)|d\lambda , &\int |P_{x'y}(\lambda) - P_{x'y'}(\lambda)|d\lambda \leq M_2 ;\\ P_{xy}(\lambda) = P_{x'y}(\lambda) \,&\text{and} \, P_{xy'}(\lambda)= P_{x'y'}(\lambda).
 \end{split}
 \end{equation}
 Hence $J(\lambda)$ takes the form,
 \begin{equation}\label{B4i}
 \begin{split}
    J(\lambda)&= P_{xy}(\lambda)|m_{1} - n_{1}| + P_{xy'}(\lambda) |m_1 - n_2|\\
     &+ P_{xy}(\lambda)|m_3 - n_3| + P_{xy'}(\lambda)|m_3 + n_4 -1|.
 \end{split}
\end{equation}

The left hand side of the Eq. (\ref{B1}) is obtained by integrating both sides of Eq. (\ref{B3}) over $\lambda$ and using Eq.(\ref{B4}), hence
\begin{equation}\label{S12}
\begin{split}
   &\langle xy \rangle + \langle xy' \rangle + \langle x'y \rangle - \langle x'y' \rangle\\
    &= \int T(\lambda)d\lambda \leq 4 - 2 \int J(\lambda)d\lambda.
 \end{split}
\end{equation}
In order to get the maximum value of the left hand side of Eq. (\ref{S12}), we have to minimize the integral of the positive quantity $J(\lambda)$ under the restriction given in the statement of the theorem. There are four possible cases corresponding to the four possible values of $(m_1,\,m_3)$: $(0$,\,0)$, $(1$,\,0)$, $(0$,\,1)$ \, \text{and} \, $(1$,\,1)$. Similarly(as in the $M_2 = 0$ case), one trivially has
\begin{equation}\label{B12i}
   J(\lambda)\geq (1-S_{1\rightarrow2})\min\{P_{xy},\,P_{xy'}\}.
\end{equation}
Then Eq. (\ref{S12}) becomes (by using (\ref{B12i})),
 \begin{equation}\label{B14}
  \begin{split}
   &\langle xy \rangle + \langle xy' \rangle + \langle x'y \rangle - \langle x'y' \rangle \\
   &\leq 4 - 2(1-S_{1\rightarrow2})\int \min\{P_{xy},\,P_{xy'}\} d\lambda.
  \end{split}
 \end{equation}
 Now,
 \begin{equation}\label{B15}
 \min\{P_{xy},\,P_{xy'}\}\,=\,\frac {1}{2}( P_{xy} + P_{xy'}) - \frac{1}{2}|P_{xy} - P_{xy'}|.
 \end{equation}
Hence,
 \begin{equation}\label{B16}
  \int \min\{P_{xy},P_{xy'}\} d\lambda\, \geq \, \max\{0,1 - \frac{M_2}{2}\}.
 \end{equation}
Then, one obtains the bound(using Eq.(\ref{B14}) and Eq.(\ref{B16})),

\hspace{1.5cm}$\langle xy \rangle + \langle xy' \rangle + \langle x'y \rangle$
\begin{equation}\label{B17}
   - \langle x'y' \rangle
  \leq 4 - 2(1-S_{1\rightarrow2})\max\{0,1 - \frac{M_2}{2}\}
\end{equation}
Now, just as in the previous case (i.e for $M_2 = 0$), it is clear that
\begin{equation}\label{B18}
\begin{split}
   B(I_2, S_{1\rightarrow 2}, M_2) &= 4 - (1 - I_2)(2 - M_2),\,\textmd{for} \, S_{1\rightarrow 2} < 1 - 2I_2 \\
    &\hspace{2cm}\textmd{and} \, M_2 < 2 \,(\text{tight upper bound})\\
    &= 4 - (1 - S_{1\rightarrow2})(2 - M_2),\,\textmd{for} \, S_{1\rightarrow 2} \geq 1 - 2I_2\\
    &\hspace{2cm}\textmd{and} \, M_2 < 2 \,(\text{tight upper bound})\\
                &= 4,                    \qquad \qquad \qquad \qquad \qquad \textmd{otherwise}.
\end{split}
\end{equation}
It still remain to show that the bound in Eq.(\ref{B16}) is tight. For this we consider a model with two variables $\lambda_1$ and $\lambda_2$ in Table 1 similar to that given in \cite{Manik}. It is clear from the table that $\int \min\{P_{xy},P_{xy'}\} d\lambda\, = \, \max\{0,1 - \frac{M_2}{2}\}$. \qquad\qquad \qquad\qquad\qquad\qquad\qquad\qquad\qquad\qquad\qquad\qquad $\blacksquare$\\
\begin{table}
  \centering
\caption{A class of deterministic no signaling model.}
\begin{tabular}{|c|c|c|c|c|c|c|c|c|}
 \hline
  $\lambda$   & $x(\lambda)$ & $x'(\lambda)$ & $y(\lambda)$ & $y'(\lambda)$ & $P_{xy}(\lambda)$ & $P_{xy'}(\lambda)$ & $P_{x'y}(\lambda)$ & $P_{x'y'}(\lambda)$ \\
 \hline
  $\lambda_1$ & -a         & a           & -a         & -a          & 0               & p                & 0                & p \\
  $\lambda_2$ &  b         & b           &  b         &  b          & 1               & 1-p              & 1                & 1-p \\
  \hline
\end{tabular}
\end{table}
\end{document}